%% file: paper.tex
\documentclass[a4paper,12pt]{amsart}

\usepackage{graphicx}
\usepackage{amsmath}
\usepackage{amssymb}
\usepackage{feynmp}
\usepackage{listings}
\usepackage{subfigure}
\usepackage{hyperref}

\lstnewenvironment{code}
    {\lstset{%
        basicstyle=\small\ttfamily,
        escapechar={\#}, 
        columns=fullflexible
    }}
    {}
\newcommand{\opensquare}{\mbox{$\rlap{$\sqcap$}\sqcup$}}
\newcommand{\fullsquare}{\,\vrule height5pt depth0pt width5pt}
\newcommand{\opencircle}{\mbox{\Large$\circ\,$}}  % moved Large outside maths

\begin{document}
\title{Computations and generation of elements on the Hopf algebra of Feynman graphs}

\author{Michael Borinsky}

\address{Humboldt-University Berlin \\Departments of Physics and Mathematics\\Unter den Linden 6\\10099 Berlin}

\email{borinsky@physik.hu-berlin.de}

\begin{abstract}
Two programs, feyngen and feyncop, were developed. feyngen is designed to generate high loop order Feynman graphs for Yang-Mills, QED and $\phi^k$ theories. feyncop can compute the coproduct of these graphs on the underlying Hopf algebra of Feynman graphs. The programs can be validated by exploiting zero dimensional field theory combinatorics and identities on the Hopf algebra which follow from the renormalizability of the theories. A benchmark for both programs was made. 
\end{abstract}

\maketitle

\section{Introduction}
The Hopf algebra structure of Feynman graphs has been explored extensively in the last years. It proved to be valuable for the analytic computation of Feynman amplitudes by means of systematic parametric integration techniques and could lead to new non-perturbative results in the scope of Dyson-Schwinger equations. \textbf{feyngen} and \textbf{feyncop} were developed to provide input for the powerful new techniques. \textbf{feyngen} is a tool for the fast generation of higher loop Feynman diagrams. \textbf{feyncop} can be used to calculate the coproduct on the Hopf algebra of Feynman graphs. This coproduct encodes the BPHZ algorithm necessary to evaluate the finite amplitude of a Feynman diagram and fits well into the world of Dyson-Schwinger equations. In this framework certain identities can be obtained which were used to validate the two programs. 

\section{Feynman diagram generation with \textbf{feyngen}}
The \textbf{python} program \textbf{feyngen} can generate $\phi^k$ for $k \geq 3$, QED, QED with Furry's theorem, Yang-Mills and $\phi^3 + \phi^4$ diagrams ready to be used in green's function calculations. Developing \textbf{feyngen}, the focus was on the generation of Feynman diagrams with comparatively large loop orders. Additionally to the generation of non-isomorphic diagrams, \textbf{feyngen} calculates the symmetry factors of the resulting graphs. Handling of graphs with fixed external legs and without is supported. Furthermore, options are available to filter for connected, one-particle-irreducible (1PI), vertex-2-connected and snail free graphs. To achieve the high speed for the computation \textbf{feyngen} relies on the established \textbf{nauty} package \cite{McKay201494,mckay1998isomorph}. The output of \textbf{feyngen} is designed to be readable by a \textbf{maple} program.

Details to the implementation, theoretical background and handling are given in \cite{Borinsky2014}.

\subsection{Examples}
Consider the sum of all two loop, photon propagator residue type, 1PI, QED diagrams. For convenience the vertices of the graph in the illustration are labeled as in the output of \textbf{feyngen}. The labels do not have further meaning. Note that, also the external source vertices are labeled, because they also appear in the output of \textbf{feyngen}.
\begin{align*}
   &   \parbox[c][30pt][t]{55pt}{\centering
   \begin{fmffile}{example9}
    \begin{fmfgraph*}(50,25)
\end{fmfgraph*}
    \end{fmffile}} + 
   \parbox[c][30pt][t]{55pt}{\centering
   \begin{fmffile}{example10}
    \begin{fmfgraph*}(50,25)
\end{fmfgraph*}
    \end{fmffile}} +
    \parbox[c][30pt][t]{55pt}{\centering
   \begin{fmffile}{example8}
    \begin{fmfgraph*}(50,25)
\end{fmfgraph*}
    \end{fmffile}}.
\end{align*}

\textbf{feyngen} generates them if it is called with the command line
\begin{code}
#\$# ./feyngen --qed 2 -b2 -p
qed_f0_b2_h2 :=
+G[[0,1,f],[1,2,f],[2,3,f],[3,0,f],[3,2,A],[4,0,A],[5,1,A]]/1
+G[[0,1,f],[1,2,f],[2,3,f],[3,0,f],[2,1,A],[4,0,A],[5,3,A]]/1
+G[[0,3,f],[1,2,f],[2,0,f],[3,1,f],[3,2,A],[4,0,A],[5,1,A]]/1
;
\end{code}
\textbf{-{}-qed} indicates QED graph generation, \textbf{2} 
stands for $2$-loop diagrams ($\hbar^2$), \textbf{-b2} makes \textbf{feyngen} 
generate graphs with 2 photon legs and 
the \textbf{-p} option filters out non 1PI graphs.

For the sum of all one loop, gauge boson propagator residue type, 1PI, Yang-Mills diagrams, 
\begin{align*}
   &   \parbox[c][30pt][t]{55pt}{\centering
   \begin{fmffile}{ym_example_1pi_fpg}
    \begin{fmfgraph*}(50,25)
\end{fmfgraph*}
    \end{fmffile}} + 
    \parbox[c][30pt][t]{55pt}{\centering
   \begin{fmffile}{ym_example_1pi_gb}
    \begin{fmfgraph*}(50,25)
\end{fmfgraph*}
    \end{fmffile}} + 
    \parbox[c][30pt][t]{55pt}{\centering
   \begin{fmffile}{ym_example_1pi_f}
    \begin{fmfgraph*}(50,25)
\end{fmfgraph*}
    \end{fmffile}} .
\end{align*}

the call, 
\begin{code}
#\$# ./feyngen --ym 1 -tp -b2 
\end{code}
where the generation of Yang-Mills graphs is triggered with the  \textbf{-{}-ym} option,
gives the desired result:
\begin{code}
ym_f0_g0_b2_h1 :=
+G[[0,1,c],[1,0,c],[2,0,A],[3,1,A]]/1
+G[[0,1,f],[1,0,f],[2,0,A],[3,1,A]]/1
+G[[1,0,A],[1,0,A],[2,0,A],[3,1,A]]/2
;
\end{code}

\subsection{Validation}
To validate the Feynman graph generation with the program \textbf{feyngen}, the perturbation expansion of a zero dimensional quantum field theory was used. 
Given for instance the generating function for $\phi^k$ theory in zero dimensions:
\begin{align}
    \label{eqn_Z_generating}
    Z_{\phi^k}( \lambda, j ) &:= 
    \int \limits_\mathbb{R} \frac{d \phi}{\sqrt{2 \pi}} ~
    e^{ - \frac{\phi^2}{2} + \lambda \frac{\phi^k}{k!} + j \phi },
\end{align}
a powerseries expansion in terms of the coupling $\lambda$ can be readily obtained,
\begin{align}
\label{eqn_Z_sum}
\widetilde{Z}_{\phi^k}( \lambda, j ) &= 
\sum \limits_{l \ge 0} 
\sum \limits_{ \substack{ n,m \ge 0 \\n k + m = 2 l } }
\frac{(2l -1)!!}{n! m! (k!)^n}  \lambda^n j^m.
\end{align}
With this multivariate powerseries the sum of the symmetry factors of disconnected $\phi^k$ with a fixed number of vertices and external legs can be obtained. Here, $\lambda$ counts the number of vertices and $j$ the number of external legs. 
The corresponding powerseries for the connected diagrams can be calculated by taking the logarithm. The reason for this is that Feynman diagrams are a labeled combinatorial class for which the exponential theorem holds \cite{flajolet2009analytic}:
\begin{align}
W(\lambda, j) &= \log( Z(\lambda,j) ).
\end{align}
For the computation of the numbers for 1PI diagrams the classical field, 
\begin{align}
\phi_c(\lambda,j) := \frac{\partial W}{\partial j},
\end{align}
is needed. The source variable $j\rightarrow j'+j_0$ is redefined such that $\phi_c(j')$ vanishes at $j'=0$. 
Using the definition of the effective action as Legendre transformation of $W$, changing $j'$ for $\phi_c$, 
\begin{align}
\Gamma = W - j' \phi_c,
\end{align}
the sum of the symmetry factors of the 1PI diagrams can be calculated using the Lagrange inversion theorem \cite{flajolet2009analytic}:
\begin{align}
  \left[ \phi_c^m \right] \Gamma(\lambda, \phi_c) = -\frac{1}{m} \left[ {j'}^{(m-2)} \right] \frac{\partial^2 W(\lambda,j')}{\partial {j'}^2} \left( \frac{j'}{\phi_c(\lambda,j')}\right)^m,
\end{align}
where $[ \cdot ]$ is the coefficient extraction operator. $\Gamma(\lambda, \phi_c)$ generates the proper green's functions in zero dimensions.
\subsection{Benchmarks}
\begin{figure}
  \subfigure[Time to generate all 1PI diagrams of given loop order.]{
  \begin{minipage}[b]{.55\linewidth}
  \centering
  \input{plot0}
  \end{minipage}}%
  \quad
  \subfigure[Average generation time for one diagram for the given loop order.]{
  \begin{minipage}[b]{.55\linewidth}
  \centering
  \input{plot0_rel}
  \end{minipage}}
  \caption{Plot of the results of the benchmark for \textbf{feyngen}. \textbf{Legend:}
  $+$ : $\phi^4$ proper propagator, 
  $\times$ : $\phi^4$ proper vertex,
  $*$ : $\phi^3$ proper propagator,
  $\opensquare$ : $\phi^3$ proper vertex,
  $\fullsquare$ : QED proper photon propagator, 
  $\opencircle$ : QCD proper gluon propagator.}
  \label{fig:1}
\end{figure}
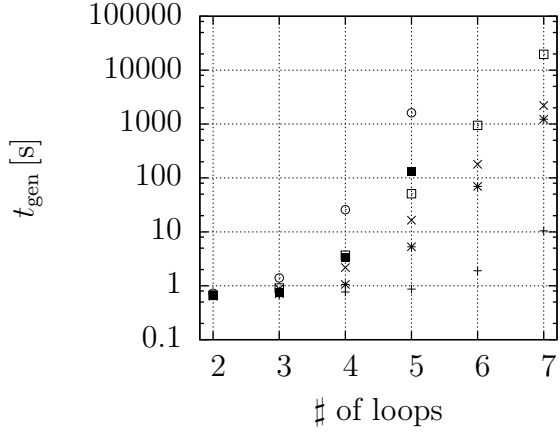
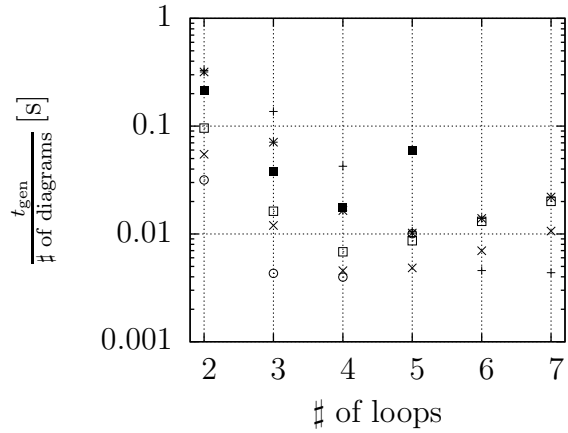
Figure \ref{fig:1} depicts an example for the computation time for the 1PI diagrams generation of a given loop order. Additionally to the non-isomorphic diagrams the corresponding symmetry factors was computed. The benchmark was performed on a \texttt{Intel(R) Core(TM) i7-3770 CPU @ 3.40GHz}. Although \textbf{feyngen} does not explicitly use parallelization, a speedup was gained because the generation of graphs using \textbf{nauty} runs in parallel to the refinement of the graphs to Feynman diagrams. 

The benchmark clearly shows that the $\phi^4$-diagram generation is the fastest. The generation of these diagrams was the main purpose for the development of \textbf{feyngen}. Therefore, the highest loop orders can be achieved in this theory. 
The exponential rise in the computation time for the QED diagram generation can be explained by the very naive diagram refinement algorithm used. The same explosion in computation time can be expected for higher loop order QCD diagrams, where the same simple algorithm was applied. 

\section{The Hopf algebra of Feynman graphs}
As was shown by Kreimer et al. \cite{connes2000renormalization,Kreimer1999_2,Kreimer1999} a Hopf algebra $\mathcal{H}_D$ can be used to describe the self-similar structure of Feynman graphs and their renormalization. The index $D$ stands for the dimension of spacetime. The coproduct $\Delta_D$ on this Hopf algebra corresponds to the forest formula in BPHZ renormalization \cite{bogoliubov1957multiplication,hepp1966proof,zimmermann1969}.
For 1PI graphs $\Gamma$ the coproduct is defined as, 
\begin{align}
\label{eqn:def_cop}
    &\Delta_D \Gamma := 
    \sum \limits_{ \substack{ \gamma \unlhd \Gamma} }
        \gamma \otimes \Gamma/\gamma& &:& &\mathcal{T} \rightarrow \mathcal{H}_D \otimes \mathcal{H}_D
\end{align}
where $\mathcal{T}$ is the set of all 1PI graphs and
\begin{align}
\label{eqn:relation_unlhd}
\gamma \unlhd \Gamma &\Leftrightarrow \gamma \in 
\left\{ \delta\subseteq \Gamma \left| \delta = 
\bigcup \limits_i \delta_i \text{, \normalfont{such that} }
\delta_i \in \mathcal{T} \right. 
\text{ \normalfont{and} } \omega_D(\delta_i) \le 0 \right\} 
\end{align}
denotes the membership of $\gamma$ in the set of subgraphs of $\Gamma$, whose connected components are superficially divergent 1PI graphs.  Disconnected graphs $\gamma=\bigcup \limits_i \gamma_i$ are identified with the product $\left(\prod \limits_i \gamma_i\right) \in  \mathcal{H}_D$.  $\Gamma/\gamma$ denotes the contraction of the subgraph $\gamma$ in $\Gamma$. 
The cograph $\Gamma/\Gamma$ and the empty graph $\gamma = \emptyset$ in \eqref{eqn:def_cop} are identified with the unit $\mathbb{I} \in \mathcal{H}_D$.

The function $\omega_D: \mathcal{T} \rightarrow \mathbb{Z}$ has an important role. It assigns the superficial degree of divergence in $D$ dimensions to a 1PI Feynman graph. $\omega_D$ performs power counting on a graph in the sense of Weinberg's theorem \cite{Weinberg1960}.

Additionally, the reduced coproduct $\widetilde{\Delta}_D$ is defined as
\begin{align}
\label{eqn:def_red_cop}
    \widetilde{\Delta}_D := \Delta_D - \text{id} \otimes \mathbb{I} - \mathbb{I} \otimes \text{id}
    & &:& &\mathcal{H}_D \rightarrow \mathcal{H}_D \otimes \mathcal{H}_D, 
\end{align}
giving rise to the space of primitive elements of $\mathcal{H}_D$:
\begin{align}
    \text{Prim}\left(\mathcal{H}_D\right) := \text{ker } \widetilde{\Delta}.
\end{align}
The primitive 1PI graphs are also called skeleton graphs.

Details to the Hopf algebra of Feynman graphs in the scope of the coproduct calculation with \textbf{feyncop} are given in \cite{Borinsky2014}.

\section{Coproduct computation with \textbf{feyncop}}
The \textbf{python} program \textbf{feyncop} can be used to compute the reduced coproduct $\widetilde{\Delta}_D$ of given 1PI graphs as defined in \eqref{eqn:def_red_cop}.  The output of \textbf{feyngen} can be piped into \textbf{feyncop} to calculate the reduced coproduct of all 1PI graphs of a given loop order and residue type.

By default, the subgraphs composed of superficially divergent, 1PI graphs of the input graphs are computed and given as output.  These correspond to the left-hand factor of the tensor product originating from the coproduct.  Optionally, the complementary cographs, giving account to the right-hand factor of the tensor product, can be computed.  Furthermore, there is the option to identify the  sub- and cographs with unlabeled 1PI graphs i.e. elements of $\mathcal{H}_D$.
Additionally, the input graphs can be filtered for primitive graphs.

The coproduct calculation does only take the degree of divergence obtained by power counting, formulated by the map $\omega_D$ into account. Further information, as gained by Furry's theorem in the case of QED, is not used.

\subsection{Examples}
The graph 
$\parbox[c][20pt][t]{22.5pt}{\centering
    \begin{fmffile}{doubleeye_unlabeled}
    \begin{fmfgraph*}(20,20)
\end{fmfgraph*}
    \end{fmffile}
    }$ is represented as an edge list using an auxiliary 
vertex labeling, as in the output of \textbf{feyngen}:
\begin{code}
G[[1,0],[2,0],[2,0],[3,1],[3,1],[3,2],[4,0],[5,1],[6,2],[7,3]].
\end{code}
This can be used as input for \textbf{feyncop}:
\begin{code}
#\$# echo "G[[1,0],[2,0],[2,0],[3,1],[3,1],[3,2],[4,0],
    [5,1],[6,2],[7,3]]" | ./feyncop -D4
\end{code}
This will yield the output:
\begin{code}
+ D[G[[1,0],[2,0],[2,0],[3,1],[3,1],
    [3,2],[4,0],[5,1],[6,2],[7,3]],
[{{1,2}}, {{3,4}}, {{1,2},{3,4}}]]
;
\end{code}
The output line 
\begin{code}
[{{1,2}}, {{3,4}}, {{1,2},{3,4}}]
\end{code}
corresponds to the subgraphs which are composed of superficially divergent, 1PI graphs, represented a by their edge sets. The edges are indexed by their order of appearance in the edge list. 
\begin{align*}
&\parbox[c][50pt][t]{65pt}{\centering
    \begin{fmffile}{doubleeye_unlabeled_subgraph1}
    \begin{fmfgraph*}(50,50)
\end{fmfgraph*}
    \end{fmffile}
    },&
&\parbox[c][50pt][t]{65pt}{\centering
    \begin{fmffile}{doubleeye_unlabeled_subgraph2}
    \begin{fmfgraph*}(50,50)
\end{fmfgraph*}
    \end{fmffile}
    }& &\text{and} &
&\parbox[c][50pt][t]{65pt}{\centering
    \begin{fmffile}{doubleeye_unlabeled_subgraph3}
    \begin{fmfgraph*}(50,50)
\end{fmfgraph*}
    \end{fmffile}
    },
\intertext{represented as the sets of sets, }
&\texttt{\{\{1,2\}\}},& &\texttt{\{\{3,4\}\}}& &\text{and}& &\texttt{\{\{1,2\},\{3,4\}\}}.
\end{align*}

\textbf{feyncop} can also be used to identify the subgraphs with elements of the Hopf algebra of Feynman graphs.
Giving again $\parbox[c][20pt][t]{22.5pt}{\centering
    \begin{fmffile}{doubleeye_unlabeled}
    \begin{fmfgraph*}(20,20)
\end{fmfgraph*}
    \end{fmffile}
    }$ as input:
\begin{code}
#\$# echo "G[[1,0],[2,0],[2,0],[3,1],[3,1],[3,2],
    [4,0],[5,1],[6,2],[7,3]]" | ./feyncop -D4 -u
\end{code}
The output will be:
\begin{code}
+ 2/1 * T[ G[[1,0],[1,0],[2,0],[3,0],[4,1],[5,1]],
  G[[1,0],[1,0],[2,0],[2,1],[3,2],[4,2],[5,0],[6,1]] ]
+ T[ (G[[1,0],[1,0],[2,0],[3,0],[4,1],[5,1]])^2,
  G[[1,0],[1,0],[2,0],[3,0],[4,1],[5,1]] ]
;
\end{code}
This output corresponds to the tensor products on the right-hand side of 
\begin{align*}
\widetilde\Delta_4 \left(
    \parbox[c][20pt][t]{22.5pt}{\centering
    \begin{fmffile}{doubleeye_mini_cop1}
    \begin{fmfgraph}(20,20)
\end{fmfgraph}
    \end{fmffile}
    }\right)
   =2
    \parbox[c][20pt][t]{13.5pt}{\centering
    \begin{fmffile}{doubleeye_mini_cop2}
    \begin{fmfgraph}(10,20)
\end{fmfgraph}
    \end{fmffile}
    }
\otimes 
\parbox[c][20pt][t]{22.5pt}{\centering
    \begin{fmffile}{doubleeye_mini_cop3}
    \begin{fmfgraph}(20,20)
\end{fmfgraph}
    \end{fmffile}
    }
 +  \left(
    \parbox[c][20pt][t]{13.5pt}{\centering
    \begin{fmffile}{doubleeye_mini_cop2}
    \begin{fmfgraph}(10,20)
\end{fmfgraph}
    \end{fmffile}
    }
\right)^2
\otimes
    \parbox[c][20pt][t]{13.5pt}{\centering
    \begin{fmffile}{doubleeye_mini_cop2}
    \begin{fmfgraph}(10,20)
\end{fmfgraph}
    \end{fmffile}
    }.
\end{align*}
where the graphs are again represented using an auxiliary vertex labeling.

\subsection{Benchmarks}
\begin{figure}
  \subfigure[Time to calculate the coproduct of all 1PI diagrams of given loop order.]{
  \begin{minipage}[b]{.55\linewidth}
  \centering
  \input{plot0cop}
  \end{minipage}}%
  \quad
  \subfigure[Average calculation time for one diagram for the given loop order.]{
  \begin{minipage}[b]{.55\linewidth}
  \centering
  \input{plot0_relcop}
  \end{minipage}}
  \caption{Plot of the results of the benchmark for \textbf{feyncop}. \textbf{Legend:}
  $+$ : $\phi^4$ vertex type diagrams,
  $\times$ : $\phi^3$ vertex type diagrams,
  $*$ : QED vertex type diagrams, 
  $\opensquare$ : QCD 3-gluon vertex type diagrams.}
  \label{fig:2}
\end{figure}
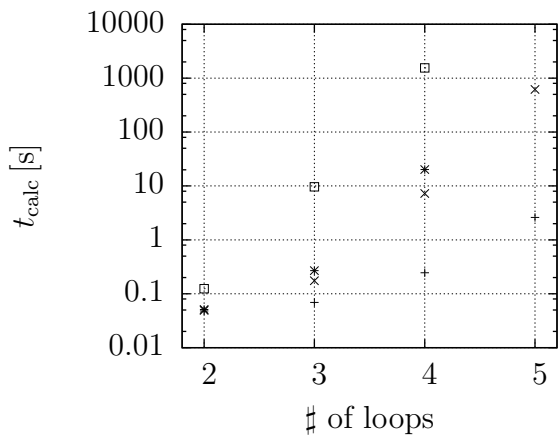
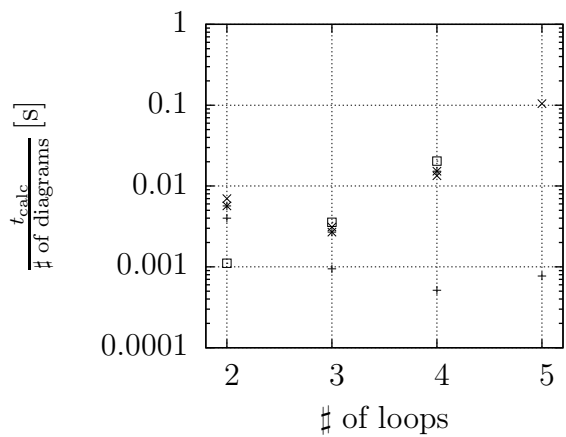
Figure \ref{fig:2} depicts an example of the computation time of the coproducts of certain classes of 1PI diagrams of a given loop order. The benchmark was performed on a \texttt{Intel(R) Core(TM) i7-3770 CPU @ 3.40GHz}. No parallelization was used. 

Because $\phi^4$-diagrams have less edges per loop in comparison to the other diagram classes, this coproduct computation is by far the fastest. There are no significant differences in the coproduct computation times of the other classes. This difference in the performance could be made much smaller by implementing a more elaborate 1PI subdiagram detection algorithm. The fast handling $\phi^4$-theory was the main priority during the development of \textbf{feyngen} and \textbf{feyncop}, so this optimization was not implemented. 

\subsection{Validation}
The output of \textbf{feyncop} can be checked by using an identity from \cite{suijlekom2007ren} on sums of Feynman graphs:
\begin{align}
\label{eqn:identity}
\sum \limits_{ \Gamma \in \mathcal{T}} \frac{\Delta_D \Gamma}{\left| \text{\normalfont{Aut}}(\Gamma) \right|} = 
\sum \limits_{\substack{\gamma = \left(\prod \limits_i \gamma_i\right) \in \mathcal{F}\\ \omega_D(\gamma_i) \leq 0}} \sum \limits_{\widetilde\Gamma \in \mathcal{T}}
\frac{ \left| \mathcal{I}( \widetilde\Gamma|\gamma ) \right| }{ 
    \left| \text{\normalfont{Aut}}( \gamma ) \right| \left| \text{\normalfont{Aut}}(\widetilde\Gamma) \right|}
    \gamma \otimes \widetilde\Gamma,
\end{align}
where $\left| \mathcal{I}( \widetilde\Gamma|\gamma ) \right|$ is the number of insertions of $\gamma$ into $\widetilde\Gamma$, $\mathcal{T}$ the set of all 1PI graphs, $\mathcal{F}$ the set of all products of 1PI graphs and $|\text{Aut}(\Gamma)|$ is the number of automorphisms of the graph $\Gamma$. 

For the validation the coproduct is calculated for the left hand side of equation \eqref{eqn:identity} and is compared to the right hand side which just depends on simple topological properties of the underlying Feynman graphs.

\section{Conclusions}
Two programs were presented. \textbf{feyngen} can generate Feynman diagrams of various theories and with certain optional properties and \textbf{feyncop} calculates the coproduct on the Hopf algebra of Feynman graphs. The method of validation and a benchmark were presented for both programs.

Both programs are publicly available at 
\\
\url{http://people.physik.hu-berlin.de/~borinsky/}.

\section*{Acknowledgements}
I wish to thank Dmitrii Batkovich for fruitful discussions and comparisons of my results with the ones obtained in the work \cite{batkovich2014graphstate}. Also, I wish to thank the organizers for the nice conference. 

\bibliographystyle{plain}
\bibliography{literature}

\end{document}

%% file: plot0.tex
% GNUPLOT: LaTeX picture with Postscript
\begingroup
  \makeatletter
  \providecommand\color[2][]{%
    \GenericError{(gnuplot) \space\space\space\@spaces}{%
      Package color not loaded in conjunction with
      terminal option `colourtext'%
    }{See the gnuplot documentation for explanation.%
    }{Either use 'blacktext' in gnuplot or load the package
      color.sty in LaTeX.}%
    \renewcommand\color[2][]{}%
  }%
  \providecommand\includegraphics[2][]{%
    \GenericError{(gnuplot) \space\space\space\@spaces}{%
      Package graphicx or graphics not loaded%
    }{See the gnuplot documentation for explanation.%
    }{The gnuplot epslatex terminal needs graphicx.sty or graphics.sty.}%
    \renewcommand\includegraphics[2][]{}%
  }%
  \providecommand\rotatebox[2]{#2}%
  \@ifundefined{ifGPcolor}{%
    \newif\ifGPcolor
    \GPcolorfalse
  }{}%
  \@ifundefined{ifGPblacktext}{%
    \newif\ifGPblacktext
    \GPblacktexttrue
  }{}%
  % define a \g@addto@macro without @ in the name:
  \let\gplgaddtomacro\g@addto@macro
  % define empty templates for all commands taking text:
  \gdef\gplbacktext{}%
  \gdef\gplfronttext{}%
  \makeatother
  \ifGPblacktext
    % no textcolor at all
    \def\colorrgb#1{}%
    \def\colorgray#1{}%
  \else
    % gray or color?
    \ifGPcolor
      \def\colorrgb#1{\color[rgb]{#1}}%
      \def\colorgray#1{\color[gray]{#1}}%
      \expandafter\def\csname LTw\endcsname{\color{white}}%
      \expandafter\def\csname LTb\endcsname{\color{black}}%
      \expandafter\def\csname LTa\endcsname{\color{black}}%
      \expandafter\def\csname LT0\endcsname{\color[rgb]{1,0,0}}%
      \expandafter\def\csname LT1\endcsname{\color[rgb]{0,1,0}}%
      \expandafter\def\csname LT2\endcsname{\color[rgb]{0,0,1}}%
      \expandafter\def\csname LT3\endcsname{\color[rgb]{1,0,1}}%
      \expandafter\def\csname LT4\endcsname{\color[rgb]{0,1,1}}%
      \expandafter\def\csname LT5\endcsname{\color[rgb]{1,1,0}}%
      \expandafter\def\csname LT6\endcsname{\color[rgb]{0,0,0}}%
      \expandafter\def\csname LT7\endcsname{\color[rgb]{1,0.3,0}}%
      \expandafter\def\csname LT8\endcsname{\color[rgb]{0.5,0.5,0.5}}%
    \else
      % gray
      \def\colorrgb#1{\color{black}}%
      \def\colorgray#1{\color[gray]{#1}}%
      \expandafter\def\csname LTw\endcsname{\color{white}}%
      \expandafter\def\csname LTb\endcsname{\color{black}}%
      \expandafter\def\csname LTa\endcsname{\color{black}}%
      \expandafter\def\csname LT0\endcsname{\color{black}}%
      \expandafter\def\csname LT1\endcsname{\color{black}}%
      \expandafter\def\csname LT2\endcsname{\color{black}}%
      \expandafter\def\csname LT3\endcsname{\color{black}}%
      \expandafter\def\csname LT4\endcsname{\color{black}}%
      \expandafter\def\csname LT5\endcsname{\color{black}}%
      \expandafter\def\csname LT6\endcsname{\color{black}}%
      \expandafter\def\csname LT7\endcsname{\color{black}}%
      \expandafter\def\csname LT8\endcsname{\color{black}}%
    \fi
  \fi
  \setlength{\unitlength}{0.0500bp}%
  \begin{picture}(4534.00,3400.00)%
    \gplgaddtomacro\gplbacktext{%
      \csname LTb\endcsname%
      \put(1342,704){\makebox(0,0)[r]{\strut{} 0.1}}%
      \csname LTb\endcsname%
      \put(1342,1109){\makebox(0,0)[r]{\strut{} 1}}%
      \csname LTb\endcsname%
      \put(1342,1514){\makebox(0,0)[r]{\strut{} 10}}%
      \csname LTb\endcsname%
      \put(1342,1920){\makebox(0,0)[r]{\strut{} 100}}%
      \csname LTb\endcsname%
      \put(1342,2325){\makebox(0,0)[r]{\strut{} 1000}}%
      \csname LTb\endcsname%
      \put(1342,2730){\makebox(0,0)[r]{\strut{} 10000}}%
      \csname LTb\endcsname%
      \put(1342,3135){\makebox(0,0)[r]{\strut{} 100000}}%
      \csname LTb\endcsname%
      \put(1573,484){\makebox(0,0){\strut{} 2}}%
      \csname LTb\endcsname%
      \put(2066,484){\makebox(0,0){\strut{} 3}}%
      \csname LTb\endcsname%
      \put(2559,484){\makebox(0,0){\strut{} 4}}%
      \csname LTb\endcsname%
      \put(3052,484){\makebox(0,0){\strut{} 5}}%
      \csname LTb\endcsname%
      \put(3545,484){\makebox(0,0){\strut{} 6}}%
      \csname LTb\endcsname%
      \put(4038,484){\makebox(0,0){\strut{} 7}}%
      \put(176,1919){\rotatebox{-270}{\makebox(0,0){\strut{}$t_\text{gen}\left[\text{s}\right]$}}}%
      \put(2805,154){\makebox(0,0){\strut{}$\sharp$ of loops}}%
    }%
    \gplgaddtomacro\gplfronttext{%
    }%
    \gplbacktext
    \put(0,0){\includegraphics{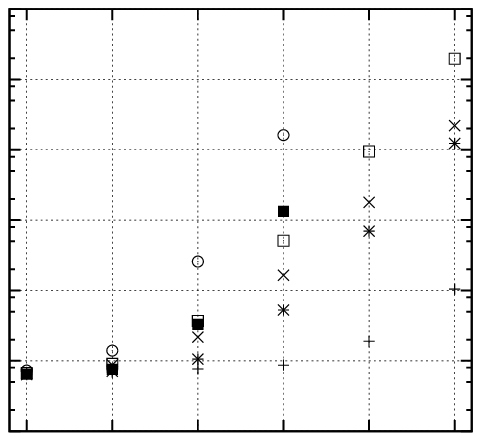}}%
    \gplfronttext
  \end{picture}%
\endgroup

%% file: plot0_rel.tex
% GNUPLOT: LaTeX picture with Postscript
\begingroup
  \makeatletter
  \providecommand\color[2][]{%
    \GenericError{(gnuplot) \space\space\space\@spaces}{%
      Package color not loaded in conjunction with
      terminal option `colourtext'%
    }{See the gnuplot documentation for explanation.%
    }{Either use 'blacktext' in gnuplot or load the package
      color.sty in LaTeX.}%
    \renewcommand\color[2][]{}%
  }%
  \providecommand\includegraphics[2][]{%
    \GenericError{(gnuplot) \space\space\space\@spaces}{%
      Package graphicx or graphics not loaded%
    }{See the gnuplot documentation for explanation.%
    }{The gnuplot epslatex terminal needs graphicx.sty or graphics.sty.}%
    \renewcommand\includegraphics[2][]{}%
  }%
  \providecommand\rotatebox[2]{#2}%
  \@ifundefined{ifGPcolor}{%
    \newif\ifGPcolor
    \GPcolorfalse
  }{}%
  \@ifundefined{ifGPblacktext}{%
    \newif\ifGPblacktext
    \GPblacktexttrue
  }{}%
  % define a \g@addto@macro without @ in the name:
  \let\gplgaddtomacro\g@addto@macro
  % define empty templates for all commands taking text:
  \gdef\gplbacktext{}%
  \gdef\gplfronttext{}%
  \makeatother
  \ifGPblacktext
    % no textcolor at all
    \def\colorrgb#1{}%
    \def\colorgray#1{}%
  \else
    % gray or color?
    \ifGPcolor
      \def\colorrgb#1{\color[rgb]{#1}}%
      \def\colorgray#1{\color[gray]{#1}}%
      \expandafter\def\csname LTw\endcsname{\color{white}}%
      \expandafter\def\csname LTb\endcsname{\color{black}}%
      \expandafter\def\csname LTa\endcsname{\color{black}}%
      \expandafter\def\csname LT0\endcsname{\color[rgb]{1,0,0}}%
      \expandafter\def\csname LT1\endcsname{\color[rgb]{0,1,0}}%
      \expandafter\def\csname LT2\endcsname{\color[rgb]{0,0,1}}%
      \expandafter\def\csname LT3\endcsname{\color[rgb]{1,0,1}}%
      \expandafter\def\csname LT4\endcsname{\color[rgb]{0,1,1}}%
      \expandafter\def\csname LT5\endcsname{\color[rgb]{1,1,0}}%
      \expandafter\def\csname LT6\endcsname{\color[rgb]{0,0,0}}%
      \expandafter\def\csname LT7\endcsname{\color[rgb]{1,0.3,0}}%
      \expandafter\def\csname LT8\endcsname{\color[rgb]{0.5,0.5,0.5}}%
    \else
      % gray
      \def\colorrgb#1{\color{black}}%
      \def\colorgray#1{\color[gray]{#1}}%
      \expandafter\def\csname LTw\endcsname{\color{white}}%
      \expandafter\def\csname LTb\endcsname{\color{black}}%
      \expandafter\def\csname LTa\endcsname{\color{black}}%
      \expandafter\def\csname LT0\endcsname{\color{black}}%
      \expandafter\def\csname LT1\endcsname{\color{black}}%
      \expandafter\def\csname LT2\endcsname{\color{black}}%
      \expandafter\def\csname LT3\endcsname{\color{black}}%
      \expandafter\def\csname LT4\endcsname{\color{black}}%
      \expandafter\def\csname LT5\endcsname{\color{black}}%
      \expandafter\def\csname LT6\endcsname{\color{black}}%
      \expandafter\def\csname LT7\endcsname{\color{black}}%
      \expandafter\def\csname LT8\endcsname{\color{black}}%
    \fi
  \fi
  \setlength{\unitlength}{0.0500bp}%
  \begin{picture}(4534.00,3400.00)%
    \gplgaddtomacro\gplbacktext{%
      \csname LTb\endcsname%
      \put(1210,704){\makebox(0,0)[r]{\strut{} 0.001}}%
      \csname LTb\endcsname%
      \put(1210,1514){\makebox(0,0)[r]{\strut{} 0.01}}%
      \csname LTb\endcsname%
      \put(1210,2325){\makebox(0,0)[r]{\strut{} 0.1}}%
      \csname LTb\endcsname%
      \put(1210,3135){\makebox(0,0)[r]{\strut{} 1}}%
      \csname LTb\endcsname%
      \put(1446,484){\makebox(0,0){\strut{} 2}}%
      \csname LTb\endcsname%
      \put(1963,484){\makebox(0,0){\strut{} 3}}%
      \csname LTb\endcsname%
      \put(2481,484){\makebox(0,0){\strut{} 4}}%
      \csname LTb\endcsname%
      \put(2998,484){\makebox(0,0){\strut{} 5}}%
      \csname LTb\endcsname%
      \put(3516,484){\makebox(0,0){\strut{} 6}}%
      \csname LTb\endcsname%
      \put(4033,484){\makebox(0,0){\strut{} 7}}%
      \put(176,1919){\rotatebox{-270}{\makebox(0,0){\strut{}$\frac{t_\text{gen}}{\sharp\text{ of diagrams}}\left[\text{s}\right]$}}}%
      \put(2739,154){\makebox(0,0){\strut{}$\sharp$ of loops}}%
    }%
    \gplgaddtomacro\gplfronttext{%
    }%
    \gplbacktext
    \put(0,0){\includegraphics{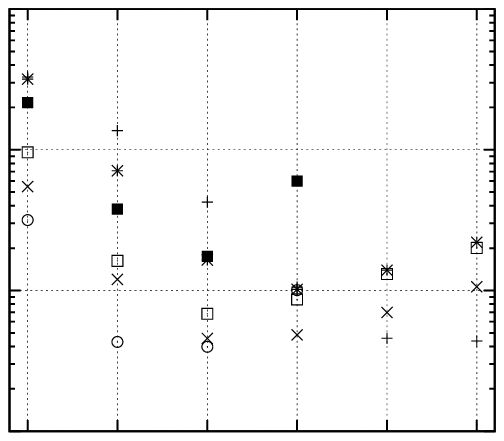}}%
    \gplfronttext
  \end{picture}%
\endgroup

%% file: plot0cop.tex
% GNUPLOT: LaTeX picture with Postscript
\begingroup
  \makeatletter
  \providecommand\color[2][]{%
    \GenericError{(gnuplot) \space\space\space\@spaces}{%
      Package color not loaded in conjunction with
      terminal option `colourtext'%
    }{See the gnuplot documentation for explanation.%
    }{Either use 'blacktext' in gnuplot or load the package
      color.sty in LaTeX.}%
    \renewcommand\color[2][]{}%
  }%
  \providecommand\includegraphics[2][]{%
    \GenericError{(gnuplot) \space\space\space\@spaces}{%
      Package graphicx or graphics not loaded%
    }{See the gnuplot documentation for explanation.%
    }{The gnuplot epslatex terminal needs graphicx.sty or graphics.sty.}%
    \renewcommand\includegraphics[2][]{}%
  }%
  \providecommand\rotatebox[2]{#2}%
  \@ifundefined{ifGPcolor}{%
    \newif\ifGPcolor
    \GPcolorfalse
  }{}%
  \@ifundefined{ifGPblacktext}{%
    \newif\ifGPblacktext
    \GPblacktexttrue
  }{}%
  % define a \g@addto@macro without @ in the name:
  \let\gplgaddtomacro\g@addto@macro
  % define empty templates for all commands taking text:
  \gdef\gplbacktext{}%
  \gdef\gplfronttext{}%
  \makeatother
  \ifGPblacktext
    % no textcolor at all
    \def\colorrgb#1{}%
    \def\colorgray#1{}%
  \else
    % gray or color?
    \ifGPcolor
      \def\colorrgb#1{\color[rgb]{#1}}%
      \def\colorgray#1{\color[gray]{#1}}%
      \expandafter\def\csname LTw\endcsname{\color{white}}%
      \expandafter\def\csname LTb\endcsname{\color{black}}%
      \expandafter\def\csname LTa\endcsname{\color{black}}%
      \expandafter\def\csname LT0\endcsname{\color[rgb]{1,0,0}}%
      \expandafter\def\csname LT1\endcsname{\color[rgb]{0,1,0}}%
      \expandafter\def\csname LT2\endcsname{\color[rgb]{0,0,1}}%
      \expandafter\def\csname LT3\endcsname{\color[rgb]{1,0,1}}%
      \expandafter\def\csname LT4\endcsname{\color[rgb]{0,1,1}}%
      \expandafter\def\csname LT5\endcsname{\color[rgb]{1,1,0}}%
      \expandafter\def\csname LT6\endcsname{\color[rgb]{0,0,0}}%
      \expandafter\def\csname LT7\endcsname{\color[rgb]{1,0.3,0}}%
      \expandafter\def\csname LT8\endcsname{\color[rgb]{0.5,0.5,0.5}}%
    \else
      % gray
      \def\colorrgb#1{\color{black}}%
      \def\colorgray#1{\color[gray]{#1}}%
      \expandafter\def\csname LTw\endcsname{\color{white}}%
      \expandafter\def\csname LTb\endcsname{\color{black}}%
      \expandafter\def\csname LTa\endcsname{\color{black}}%
      \expandafter\def\csname LT0\endcsname{\color{black}}%
      \expandafter\def\csname LT1\endcsname{\color{black}}%
      \expandafter\def\csname LT2\endcsname{\color{black}}%
      \expandafter\def\csname LT3\endcsname{\color{black}}%
      \expandafter\def\csname LT4\endcsname{\color{black}}%
      \expandafter\def\csname LT5\endcsname{\color{black}}%
      \expandafter\def\csname LT6\endcsname{\color{black}}%
      \expandafter\def\csname LT7\endcsname{\color{black}}%
      \expandafter\def\csname LT8\endcsname{\color{black}}%
    \fi
  \fi
  \setlength{\unitlength}{0.0500bp}%
  \begin{picture}(4534.00,3400.00)%
    \gplgaddtomacro\gplbacktext{%
      \csname LTb\endcsname%
      \put(1210,704){\makebox(0,0)[r]{\strut{} 0.01}}%
      \csname LTb\endcsname%
      \put(1210,1109){\makebox(0,0)[r]{\strut{} 0.1}}%
      \csname LTb\endcsname%
      \put(1210,1514){\makebox(0,0)[r]{\strut{} 1}}%
      \csname LTb\endcsname%
      \put(1210,1920){\makebox(0,0)[r]{\strut{} 10}}%
      \csname LTb\endcsname%
      \put(1210,2325){\makebox(0,0)[r]{\strut{} 100}}%
      \csname LTb\endcsname%
      \put(1210,2730){\makebox(0,0)[r]{\strut{} 1000}}%
      \csname LTb\endcsname%
      \put(1210,3135){\makebox(0,0)[r]{\strut{} 10000}}%
      \csname LTb\endcsname%
      \put(1506,484){\makebox(0,0){\strut{} 2}}%
      \csname LTb\endcsname%
      \put(2328,484){\makebox(0,0){\strut{} 3}}%
      \csname LTb\endcsname%
      \put(3151,484){\makebox(0,0){\strut{} 4}}%
      \csname LTb\endcsname%
      \put(3973,484){\makebox(0,0){\strut{} 5}}%
      \put(176,1919){\rotatebox{-270}{\makebox(0,0){\strut{}$t_\text{calc}\left[\text{s}\right]$}}}%
      \put(2739,154){\makebox(0,0){\strut{}$\sharp$ of loops}}%
    }%
    \gplgaddtomacro\gplfronttext{%
    }%
    \gplbacktext
    \put(0,0){\includegraphics{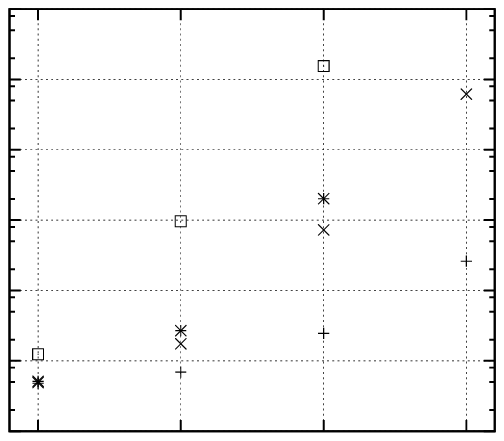}}%
    \gplfronttext
  \end{picture}%
\endgroup

%% file: plot0_relcop.tex
% GNUPLOT: LaTeX picture with Postscript
\begingroup
  \makeatletter
  \providecommand\color[2][]{%
    \GenericError{(gnuplot) \space\space\space\@spaces}{%
      Package color not loaded in conjunction with
      terminal option `colourtext'%
    }{See the gnuplot documentation for explanation.%
    }{Either use 'blacktext' in gnuplot or load the package
      color.sty in LaTeX.}%
    \renewcommand\color[2][]{}%
  }%
  \providecommand\includegraphics[2][]{%
    \GenericError{(gnuplot) \space\space\space\@spaces}{%
      Package graphicx or graphics not loaded%
    }{See the gnuplot documentation for explanation.%
    }{The gnuplot epslatex terminal needs graphicx.sty or graphics.sty.}%
    \renewcommand\includegraphics[2][]{}%
  }%
  \providecommand\rotatebox[2]{#2}%
  \@ifundefined{ifGPcolor}{%
    \newif\ifGPcolor
    \GPcolorfalse
  }{}%
  \@ifundefined{ifGPblacktext}{%
    \newif\ifGPblacktext
    \GPblacktexttrue
  }{}%
  % define a \g@addto@macro without @ in the name:
  \let\gplgaddtomacro\g@addto@macro
  % define empty templates for all commands taking text:
  \gdef\gplbacktext{}%
  \gdef\gplfronttext{}%
  \makeatother
  \ifGPblacktext
    % no textcolor at all
    \def\colorrgb#1{}%
    \def\colorgray#1{}%
  \else
    % gray or color?
    \ifGPcolor
      \def\colorrgb#1{\color[rgb]{#1}}%
      \def\colorgray#1{\color[gray]{#1}}%
      \expandafter\def\csname LTw\endcsname{\color{white}}%
      \expandafter\def\csname LTb\endcsname{\color{black}}%
      \expandafter\def\csname LTa\endcsname{\color{black}}%
      \expandafter\def\csname LT0\endcsname{\color[rgb]{1,0,0}}%
      \expandafter\def\csname LT1\endcsname{\color[rgb]{0,1,0}}%
      \expandafter\def\csname LT2\endcsname{\color[rgb]{0,0,1}}%
      \expandafter\def\csname LT3\endcsname{\color[rgb]{1,0,1}}%
      \expandafter\def\csname LT4\endcsname{\color[rgb]{0,1,1}}%
      \expandafter\def\csname LT5\endcsname{\color[rgb]{1,1,0}}%
      \expandafter\def\csname LT6\endcsname{\color[rgb]{0,0,0}}%
      \expandafter\def\csname LT7\endcsname{\color[rgb]{1,0.3,0}}%
      \expandafter\def\csname LT8\endcsname{\color[rgb]{0.5,0.5,0.5}}%
    \else
      % gray
      \def\colorrgb#1{\color{black}}%
      \def\colorgray#1{\color[gray]{#1}}%
      \expandafter\def\csname LTw\endcsname{\color{white}}%
      \expandafter\def\csname LTb\endcsname{\color{black}}%
      \expandafter\def\csname LTa\endcsname{\color{black}}%
      \expandafter\def\csname LT0\endcsname{\color{black}}%
      \expandafter\def\csname LT1\endcsname{\color{black}}%
      \expandafter\def\csname LT2\endcsname{\color{black}}%
      \expandafter\def\csname LT3\endcsname{\color{black}}%
      \expandafter\def\csname LT4\endcsname{\color{black}}%
      \expandafter\def\csname LT5\endcsname{\color{black}}%
      \expandafter\def\csname LT6\endcsname{\color{black}}%
      \expandafter\def\csname LT7\endcsname{\color{black}}%
      \expandafter\def\csname LT8\endcsname{\color{black}}%
    \fi
  \fi
  \setlength{\unitlength}{0.0500bp}%
  \begin{picture}(4534.00,3400.00)%
    \gplgaddtomacro\gplbacktext{%
      \csname LTb\endcsname%
      \put(1342,704){\makebox(0,0)[r]{\strut{} 0.0001}}%
      \csname LTb\endcsname%
      \put(1342,1312){\makebox(0,0)[r]{\strut{} 0.001}}%
      \csname LTb\endcsname%
      \put(1342,1920){\makebox(0,0)[r]{\strut{} 0.01}}%
      \csname LTb\endcsname%
      \put(1342,2527){\makebox(0,0)[r]{\strut{} 0.1}}%
      \csname LTb\endcsname%
      \put(1342,3135){\makebox(0,0)[r]{\strut{} 1}}%
      \csname LTb\endcsname%
      \put(1631,484){\makebox(0,0){\strut{} 2}}%
      \csname LTb\endcsname%
      \put(2414,484){\makebox(0,0){\strut{} 3}}%
      \csname LTb\endcsname%
      \put(3197,484){\makebox(0,0){\strut{} 4}}%
      \csname LTb\endcsname%
      \put(3980,484){\makebox(0,0){\strut{} 5}}%
      \put(176,1919){\rotatebox{-270}{\makebox(0,0){\strut{}$\frac{t_\text{calc}}{\sharp\text{ of diagrams}}\left[\text{s}\right]$}}}%
      \put(2805,154){\makebox(0,0){\strut{}$\sharp$ of loops}}%
    }%
    \gplgaddtomacro\gplfronttext{%
    }%
    \gplbacktext
    \put(0,0){\includegraphics{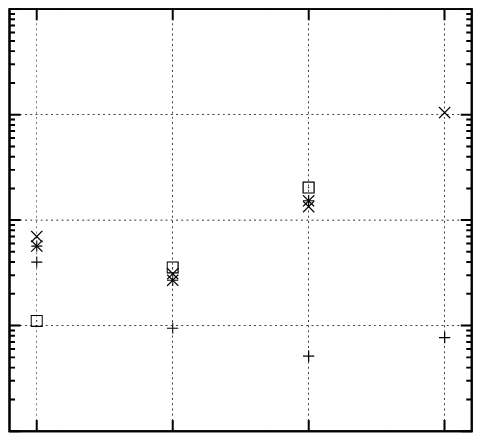}}%
    \gplfronttext
  \end{picture}%
\endgroup